# Effect of the hydrogenation solvent in the PHIP-SAH hyperpolarization of [1-$^{13}$C]pyruvate

O. Bondar,* E. Cavallari,* C.Carrera,** S. Aime,* F. Reineri*[#]

* Department of Molecular Biotechnology and Health Sciences Molecular Imaging Centre
University of Torino
Via Nizza 52, 10126 Torino (Italy)
** Institute of Biostructures and Bioimaging
National Research Council
Via Nizza 52, 10126 Torino (Italy)
[#] corresponding author email: francesca.reineri@unito.it

**Abstract**

ParaHydrogen Induced Polarization-Side Arm Hydrogenation (PHIP-SAH) is an inexpensive tool to obtain hyperpolarized pyruvate (and other metabolites) that can be applied to in-vivo diagnostics, for the investigation of metabolic processes. This method is based on hydrogenation, using hydrogen enriched in the para-isomer, of unsaturated substrates, catalyzed, usually, by a homogeneous rhodium(I) complex. In this work, the effect of the solvent on the hydrogenation efficiency and on the hyperpolarization level were investigated. Coordinating solvents, such as acetone and methanol, can increase significantly either the efficiency or the hyperpolarization level, but they are not compatible with the intended metabolic applications. The phase extraction of the hyperpolarized product (sodium pyruvate) in an aqueous solution was obtained carrying out the hydrogenation reaction in chloroform and toluene. The traces of the organic solvents in the water phase were removed, by means of filtration through a lipophilic resin, thus improving the biocompatibility of the aqueous solution of the hyperpolarized product.

## 1.1 Introduction

NMR is a non-disruptive spectroscopic method that allows detection and quantitation of small biomolecules, among which metabolites, in different matrices, in cells and in-vivo. However, the vast majority of MRI studies addressing diagnostic applications make use of the $^1$H signal of water while other molecules are not usually observed, due to their low abundance and to the intrinsic low sensitivity of MR based methods.

Hyperpolarization methods [1,2] allowed to circumvent this issue, by temporarily enhancing the MR signal of molecules by several orders of magnitude, thus allowing the detection of key molecules of diagnostic interest.

ParaHydrogen Induced Polarization (PHIP) [3,4] is a cost-effective and fast hyperpolarization technique that exploits the spin order of hydrogen enriched in the para isomer to obtain an unbalance of the spin states population of the hyperpolarization products, with respect to thermal equilibrium. In hydrogenative PHIP, both protons of a hydrogen molecule must be pairwise transferred to an unsaturated substrate. In this context, the most used catalysts are rhodium(I) complexes, that allow pairwise addition of both H$_2$ protons to the substrate molecule, through very short-lived reaction intermediates.

The *in-vivo* application of hyperpolarized substances obtained by means of PHIP has been reported since the early days of the application of hyperpolarization techniques[5,6] and recent advancements have been surveyed in both hydrogenative[7,8] and non-hydrogenative PHIP[9]. The recent achievements have definitively improved the outlook for the PHIP approach, although the gold-standard method for the diagnostic applications still remains the d-DNP modality (dissolution-Dynamic Nuclear Polarization) [10,11].

Pyruvate is the most scrutinized hyperpolarized metabolite and it has been used in the investigation of metabolism of different diseases. [12–15] Clinical trials are currently ongoing for cancer (prostate and breast cancer [16]). D-DNP can provide orders of magnitude enhanced signals on many different molecules, but it is definitively expensive and technologically demanding.

Therefore, the possibility of obtaining HP (hyperpolarized) pyruvate using PHIP-based methods is quite appealing and this task is under intense scrutiny by means of both hydrogenative [17–19] and non-hydrogenative PHIP (i.e. SABRE)[20,21] based methods.

The removal of the hydrogenation catalyst continues to be major issue for biological applications, in spite early investigations have shown relatively low toxicity[22] and its removal has been pursued by means of metal scavengers.[23] Phase transfer of the hyperpolarized product from a hydrophobic organic solvent, in which the hydrogenation is carried out, to an aqueous medium, has been exploited for both SABRE[24] and hydrogenative-PHIP[25]. This method yielded positive results in the catalyst removal, which is mostly retained in the organic phase. However, hydrophobic solvents such as chloroform are also partially soluble in water[26], thus introducing serious problems about the bio-compatibility of the hyperpolarized substrate containing solution.

In this work, our results on the effect of different reaction solvents on hyperpolarization of an unsaturated derivative of pyruvate (propargyl-pyruvate) by means of hydrogenative PHIP (PHIP-SAH) are reported. After hydrolysis of the ester and successive phase extraction, an aqueous solution of the hyperpolarized molecule was obtained and the residual amount of the organic solvents measured. Next, the aqueous solution were filtered to remove the organic solvent residues.

## 2.1 Materials and methods

### 2.1.1 Sample preparation

In order to hydrogenate the propargylic ester of pyruvate, the hydrogenation catalyst ([1,4-bis(diphenylphosphino)butane](1,5- cyclooctadiene)rhodium(I) tetrafluoroborate], Sigma Aldrich, CAS: 79255-71-3) was used, without purification.

NMR sample tubes equipped with a gas-tight valve (Norell® NMR sample tubes) were used to carry out the hydrogenation reactions.

Deuterated solvents were bought from CortecNet and used without purification. The propargylic ester of [1-$^{13}$C]pyruvate was synthesized as reported in ref [27].

The catalyst/substrates solutions were prepared and loaded into the NMR sample tubes according to the following experimental set-up:

**Acetone.** A 7mM solution of the catalyst was prepared by dissolving the commercial compound in deuterated acetone (Acetone-$d_4$). The solution was filtered through a poly(tetrafluoroethylene) (PTFE) (Whatman, UNIFLO) syringe filter, size of the pores 0.22 µm. 100 µl of this solution were transferred into each NMR sample tube and the solution was frozen in liquid nitrogen. The propargylic ester of [1-$^{13}$C]pyruvate (27 µmol in each sample tube, concentration in the hydrogenation mixture 270mM) was added to this solution.

**Chloroform.** A 14mM solution of the catalyst was prepared in deuterated chloroform (CDCl$_3$) and filtered on the PTFE filters. 100µl of this solution were transferred into each NMR sample tube and the tubes were frozen in a liquid nitrogen bath. The propargylic ester of [1-$^{13}$C]pyruvate (27 µmol) was added to each tube. In the experiments carried out using ethanol as co-solvent, 5% of EtOH (Ethanol BioUltra for molecular biology, >99.8%, Sigma-Aldrich) was added as well.

**Methanol.** Due to the low solubility of the catalyst precursor in methanol, it was dissolved in chloroform, first, and the solution was filtered through the PTFE filter (see above). The catalyst containing solution was distributed into the NMR sample tubes (1.4 µmoles/tube) and chloroform was evaporated by means of an argon flow. Deuterated methanol was added to each tube (100µl, final concentration of the catalyst 14mM) and the solution was frozen in liquid nitrogen. The propargylic ester of [1-$^{13}$C]pyruvate was added to the tubes (concentration in the reaction mixture 270mM).

**Toluene.** Due to the low solubility of the catalyst in toluene, it was dissolved in deuterated chloroform, first, and filtered through the PTFE filter (see above). The solution was then distributed into the NMR sample tubes (1µmol/tube), then chloroform was evaporated by means of an argon flow. Toluene was added (100µl toluene-$h_8$) and the tubes were frozen into a liquid nitrogen bath. The propargylic ester of [1-$^{13}$C]pyruvate (26.6 µmol) was added to the tube, as well as 5% v/v (Ethanol BioUltra for molecular biology, >99.8%, Sigma-Aldrich).

In order to add parahydrogen, the frozen NMR sample tube, loaded with the substrate/catalyst solution, was connected to a vacuum line and the air was removed by applying vacuum (<0.2 mbar). Parahydrogen (BPHG Bruker ParaHydrogen Generator 86% enrichment) (2.1 bar) was added, while keeping the NMR tube in a liquid nitrogen bath. The tube was disconnected from the vacuum line and kept frozen in liquid nitrogen (77K) in order to prevent any chemical reaction, until the start of the hyperpolarization experiment.

*2.1.2.Hyperpolarization*

In order to carry out the parahydrogenation reactions, the NMR tube containing the frozen reaction mixture was brought to RT in order to melt the solution, then immersed in a hot water bath (at 353K for 7 s), and shaken vigorously for 3 s. To obtain an efficient diffusion of hydrogen in the reaction mixture, 6-8 shaking cycles per second were applied.

### 2.1.3 $^1$H hyperpolarization

For the ALTADENA experiments, the parahydrogenation reaction was carried out in the proximity of the NMR magnet (14.1T, Bruker Avance spectrometer). Due to the fast decay of the hyperpolarized signals, magnetic field adjustments (shimming) and RF tuning cannot be performed on the solutions containing the hyperpolarized samples. Prior to the start of the hyperpolarization experiments, the NMR spectrometer is made ready by doing the shimming adjustments on a sample with similar characteristics to those ones of the HP sample (i.e. same solvent, same volume). Immediately after the end of shaking, the sample tube was opened, 250μl of solvent (the same solvent used for the reaction) were added and the NMR tube was placed into the high field NMR magnet. A single shot $^1$H-NMR spectrum was acquired (flip angle 90°, RG 1). The time delay between the end of the reaction and the acquisition of the $^1$H-NMR spectrum was 18 ± 1 s.

### 2.1.3 $^{13}$C hyperpolarization

The spin order transfer from the parahydrogen protons to $^{13}$C of the carboxylate group was obtained by applying magnetic field cycle.

The equipment for the magnetic field cycle consisted of a magnetic field shield (Bartington TLMS-C200 capped-end magnetic shield, three-layers mu-metal) containing a solenoid coil (i.d.180mm, 350mm eight) fed with electric current provided by a Keysight Arbitrary Waveform Generator (33220A 20 MHz waveform generator, Keysight Technologies, Santa Rosa, CA, U.S.A.). The electric current in the solenoid is controlled by a custom-written function (Microsoft Visual Basic) and allows a precise control of the magnetic field strength during the overall procedure. The magnetic field profile applied to all the experiments (figure S1) consists in a diabatic step from 1.5 μT (starting magnetic field) to 50nT and an adiabatic re-magnetization (exponential) to 10μT in 4 seconds. The amplitude of the field is measured by means of a three-axis fluxgate magnetometer (Mag-03, Bartington, Witney, UK).

In all the experiments, the NMR tube was placed inside the magnetic field set to 1.5μT and the MFC profile was applied.

The sample tube was removed from the shield, 250μl of solvent were added and the NMR tube was transported into the NMR spectrometer where the $^{13}$C-NMR spectrum (one shot, 90° flip angle, RG 9) was acquired.

Thermal equilibrium spectra were acquired with 8 transients, using 90° pulses separated by 250 s intervals.

*2.1.4 Hydrolysis of the ester and phase extraction*

Hydrolyses was carried out using an aqueous base solution (NaOH 0.13N) containing sodium ascorbate (50mM). For each experiment, 260 µl of this solution were charged into a PTFE tube (PTFE 1/16" x 0.75i.d. tube, VICI Jour). The tube was pressurized with Argon (1.5bar) and heated in a hot water bath (80°C). The injection of the aqueous base into the organic solution was obtained by opening the injection valve (figure S4). A few seconds after (7-8 s), an acidic buffer (HEPES 144mM, 100 µl, pH 5.4) was added to obtain neutral pH. It must be noticed that pyruvate is quite unstable in basic solution [28] and it tends to degrade quickly. In the end, the aqueous phase was collected in a syringe, transferred into an NMR sample tube for the acquisition of the $^{13}$C-NMR spectrum and diluted with 200-300 µl D$_2$O. Thermal $^{13}$C spectra were also acquired on each sample, in order to allow shorter repetition time between the scans, a Gd(III) complex (Gd-DO3A, 3mM)[29] was added to the aqueous solution, after the acquisition of the hyperpolarized signal. The paramagnetic complex allowed to shorten the relaxation time of the $^{13}$C signal and the time delay between the scans was set to 2s. The concentration of sodium pyruvate and of the other side-products in the aqueous phase were determined by adding the reference standard TSP (3-(Trimethylsilyl)propionic-2,2,3,3-d$_4$ acid CAS 24493-21-8, 15mM).

*2.1.5 Filtration*

In order to remove the organic solvents (toluene and chloroform) still present in the aqueous solution after the phase extraction, a small amount (approximately 100 µl) of TENAX resin was used. The resin was placed in a 1ml syringe and retained using a PEEK filter (Idex Fluidics, Frit in Ferrule 2 µm PEEK). The aqueous solution was put on the top of the column and quickly eluted through the resin. The water solution was collected in an NMR tube and placed into the NMR spectrometer for the acquisition of the $^{13}$C-NMR spectrum. Thermal spectra were also acquired following the addition of the Gd(III) solution (to speed up $^{13}$C-NMR acquisition) and of the standard TSP, as described in the previous paragraph. For the entire procedure, see video 1.

**3.1 Results and discussion**

*3.1.1 Hydrogenation efficiency in different solvents*

The hydrogenation reactions have been carried out in methanol, acetone, chloroform, chloroform/ethanol, toluene and toluene/ethanol.

The use of PHIP polarized substrates in diagnostic applications requires that a highly concentrated sample of the product is obtained in a very short time, in order to limit the hyperpolarization loss due to relaxation. The hydrogenation method, i.e. the use of NMR tubes pressurized with para-enriched hydrogen, allowed to use a small amount of reagents and to reach good reaction yields , in the very short time required for PHIP hyperpolarization. Different hydrogenation methods have been reported, which are based on spraying the hydrogenation mixture in a reactor filled with parahydrogen [30][31] or bubbling paraH$_2$ through the hydrogenation solvent.[32,33] The former one requires a dedicated system and a relatively large amount of reactants, whereas the latter may need high parahydrogen pressure (>10 bar)[34] to reach high efficiency of the reactions. In the herein reported experiments, a concentrated solution (240 mM) of the alkyne (propargyl-pyruvate) was hydrogenated to alkene in few seconds (1-3 seconds).

The concentration of the catalyst and the hydrogenation time were optimized for each hydrogenation solvent, in order to obtain complete hydrogenation of the substrate. The turnover frequency (TOF) of the catalyst in the different solvents was calculated (figure 1).

The highest hydrogenation efficiency was obtained in acetone, in which complete conversion of the alkyne to the alkene was obtained in only 1 second, using half of the catalyst used in the experiments carried out in methanol (7 mM instead of 14mM). Acetone and methanol are both coordinating solvents and they are widely applied for the hydrogenation reactions catalyzed by this type of rhodium complexes, containing chelating phosphines.[35] These solvents allow the formation of the catalytically active complexes (**I** in scheme 1), upon the hydrogenation of the coordinated diene that frees the coordinating sites required for binding the substrate (scheme 1). The formation of these active complexes can be detected in the $^{31}$P-NMR spectra of the hydrogenation catalyst, following to the activation (i.e. the release of the coordinated diene). In these spectra, the signals of the phosphine ligand in the activated complex, in different solvents, can be clearly distinguished from those of the catalyst precursor (Table S1).[36]

In the catalytic cycle, the so-called unsaturated route (scheme 1) [37,38], the substrate coordinates to the metal center and displaces the solvent molecules (**I** to **II** in scheme 1). The higher efficiency of the catalyst in acetone, as shown by the higher TOF, may be due to the fact that, being acetone a weaker ligand than methanol[39], it can be replaced more easily by the substrate, thus facilitating the hydrogenation reaction.

Unfortunately, acetone and methanol are completely miscible with water, therefore they are not compatible with in-vivo applications.

Vice-versa, chloroform and toluene are hydrophobic solvents, characterized by low solubility in water (toluene/water 0.052 WT% at 293K; chloroform/water 0.8 WT% at 293K)[40], therefore they can be used for the phase-extraction step of PHIP hyperpolarized products.

So far, chloroform has been used to obtain the phase-extraction of PHIP hyperpolarized products[24,25] from the organic solution to the aqueous phase. It is a non-coordinating solvent and the activation of the catalyst precursor, by means of the hydrogenation of the coordinated diene (cyclooctadiene or norbornadiene), leads to the formation of non-catalytically active dimers.[36] Nevertheless, when the unsaturated substrate is added to the reaction mixture before the activation of the catalyst, the efficiency of the hydrogenation reaction is the same as the one observed in methanol (figure 1). The observed behavior can be accounted for considering that, in the hydrogenation pathway (scheme 1), the substrate can replace the product molecule directly (passage **V** to **II**, dotted arrow in scheme 1) and the formation of the solvated specie (form **I**, scheme 1) can be circumvented. The addition of a coordinating solvent (i.e. ethanol) does not improve the reaction efficiency, although the hyperpolarization level is increased, as it will be shown in the following section.

Toluene can coordinate to the metal center, as it can be observed from the $^{31}$P-NMR signals (Table S1), but unfortunately the hydrogenation efficiency of this complex is low, probably due to its scarce solubility in this solvent. The addition of a small percentage of ethanol (5% v/v) leads to an increase in both solubility and hydrogenation efficiency. It must be outlined that ethanol has been preferred to methanol, as co-solvent, due to its lower toxicity and better bio-compatibility.

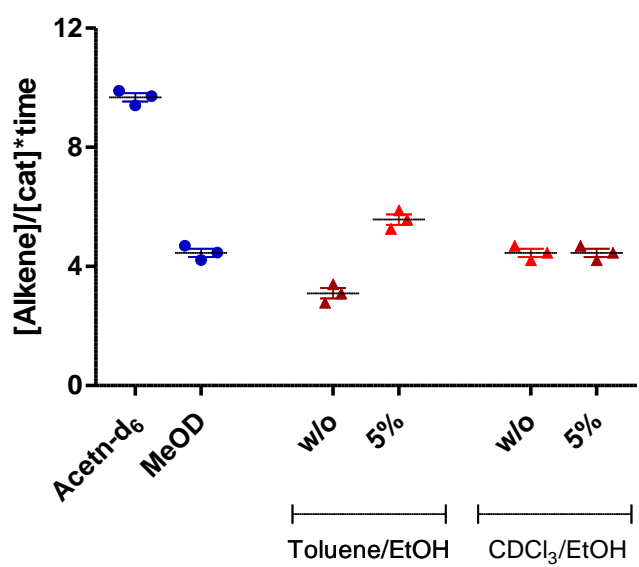

**Figure 1**. TOF (turnover frequency) (moles of substrate that a mole of catalyst can convert in a unit time, 1s) for the catalyst (Rh(I) complex containing the chelating phosphine dppb (bis-(diphenilphosphino)butane)) in different solvents. The experimental conditions (hydrogen pressure, reaction temperature) and the hydrogenation method are described in the methods section.

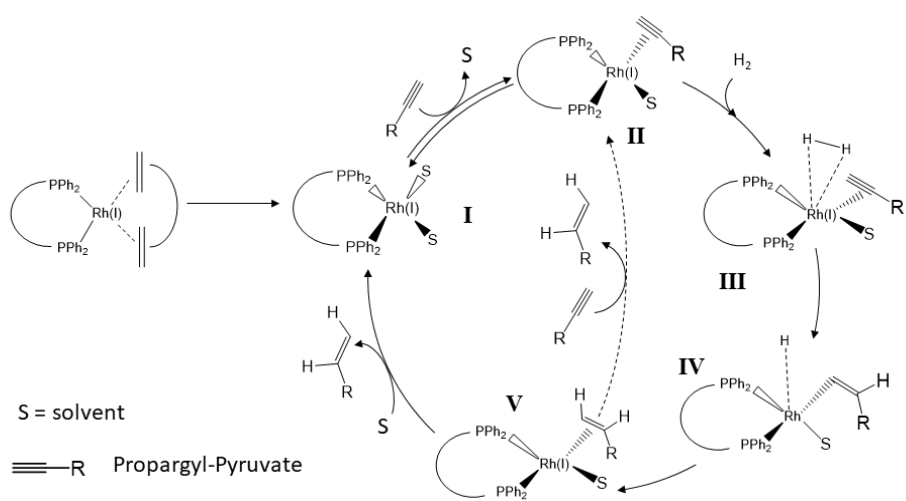

**Scheme 1**. Hydrogenation pathway for the Rh(I) complex containing the chelating phosphine dppb (bis-(diphenilphosphino)butane). In the passage shown by the dotted line, the intermediate **I** is circumvented, i.e. the adduct of catalyst and solvent is not formed. This is the case of chloroform, that is a non-coordinating solvent.

### 3.1.2 $^1H$ and $^{13}C$ Hyperpolarization on reaction products

Early on in the study of the PHIP phenomenon, it was realized that the hyperpolarized spectra of the product molecule depend crucially on the catalyst,[41] because mixing between the singlet state of parahydrogen and the other states (triplet states) occurs on hydrogenation intermediates.[42] The singlet/triplet mixing factor can be substantially different depending on the substrate, on the reaction solvent and other experimental conditions. Higher S/T mixing leads substantially to loss of singlet order and hence lower hyperpolarization level on the products. More stable reaction intermediates would lead to longer contact time of the hydrogen molecule on these intermediates and less efficient transfer of the singlet state to the products. In other words, parahydrogen is not transferred to the product in pure singlet state, thus making the hydrogenation intermediates to have an effect on the S/T mixing and, in turn, introducing a relevant weight of the experimental conditions in the determination of the polarization level on products.

In order to estimate the efficiency of the singlet state transfer to the product, $^1H$ hyperpolarized spectra have been acquired, in the different solvents. In these experiments, hyperpolarization on protons has been taken as a readout of the singlet state transferred to the product. This can be considered a more direct readout than the $^{13}C$ polarization, which reports on the efficiency of spin order transfer techniques, i.e. RF pulses or magnetic field cycle.

In these experiments, called ALTADENA,[43] the molecule is hydrogenated at geomagnetic field, with all the protons resonating at the same Larmor frequency, and then transported into the high field of the NMR spectrometer for detection. During this passage, net magnetization on the $^1$H-NMR signals of the product takes place.

The $^1$H-NMR spectrum (figure 2) obtained from the ALTADENA experiments shows that hyperpolarization is distributed on all the protons of the allyl moiety.

Despite the higher catalyst efficiency observed in acetone, proton hyperpolarization in this solvent is significantly lower than in methanol. This can be explained considering the final step of the hydrogenation pathway (step **V** to **I** scheme 1), in which the product is replaced by a solvent molecule, to allow a new molecule of the substrate to bind at the catalyst and start a new catalysis turnover. When the binding affinity of the solvent is higher, the displacement of the product molecule is more energetically favored and this step is faster. This would result in a smaller singlet/triplet mixing on the intermediate **V** (scheme 1) and higher polarization on the product.

Next, $^{13}$C hyperpolarization has been obtained (figure 2), upon the application of the magnetic field cycle procedure. Consistently with the ALTADENA results, the $^{13}$C hyperpolarization level was higher in methanol than in acetone. It must be noticed that the hyperpolarization level is different on the two isoforms of pyruvate-ester, being significantly higher on the oxo-form (2-oxo-propanoate, 17.5±0.9 % $^{13}$C hyperpolarization), than on the ketalic-form (2,2-dimetoxy-propanoate, 6.4±0.2% $^{13}$C hyperpolarization). The two forms of pyruvate are 15% (2-oxo-propanoate) and 85% (2,2-dimetoxy-propanoate) of the total amount of pyruvate, respectively.

The difference between the $^{13}$C hyperpolarization of the two forms derives from the differences observed in the polarization of the $^1$H signals. In fact, as shown in figure 2, the methylene protons appear more polarized on oxo-propanoate than on dimetoxy-propanoate.

In pure toluene, hyperpolarization on $^{13}$C is significantly lower than in the other solvents, but if a small percentage of ethanol is added (ethanol, 5% v/v), hyperpolarization is significantly increased. This can be due to the improved solubility of the catalyst in the toluene/ethanol mixture, but also to the fact that the two solvents can mutually exchange in the coordination sphere.

Chloroform is non-coordinating and it does not intervene directly in the reaction pathway. The observed hyperpolarization, on both $^1$H and $^{13}$C spectra, is slightly lower than in toluene. Also in this case, the addition of the co-solvent (ethanol, 5%) leads to an increase of the hyperpolarization of the $^1$H and $^{13}$C signals, although the effect of the co-solvent on hyperpolarization is lower than that observed in toluene (figure 3).

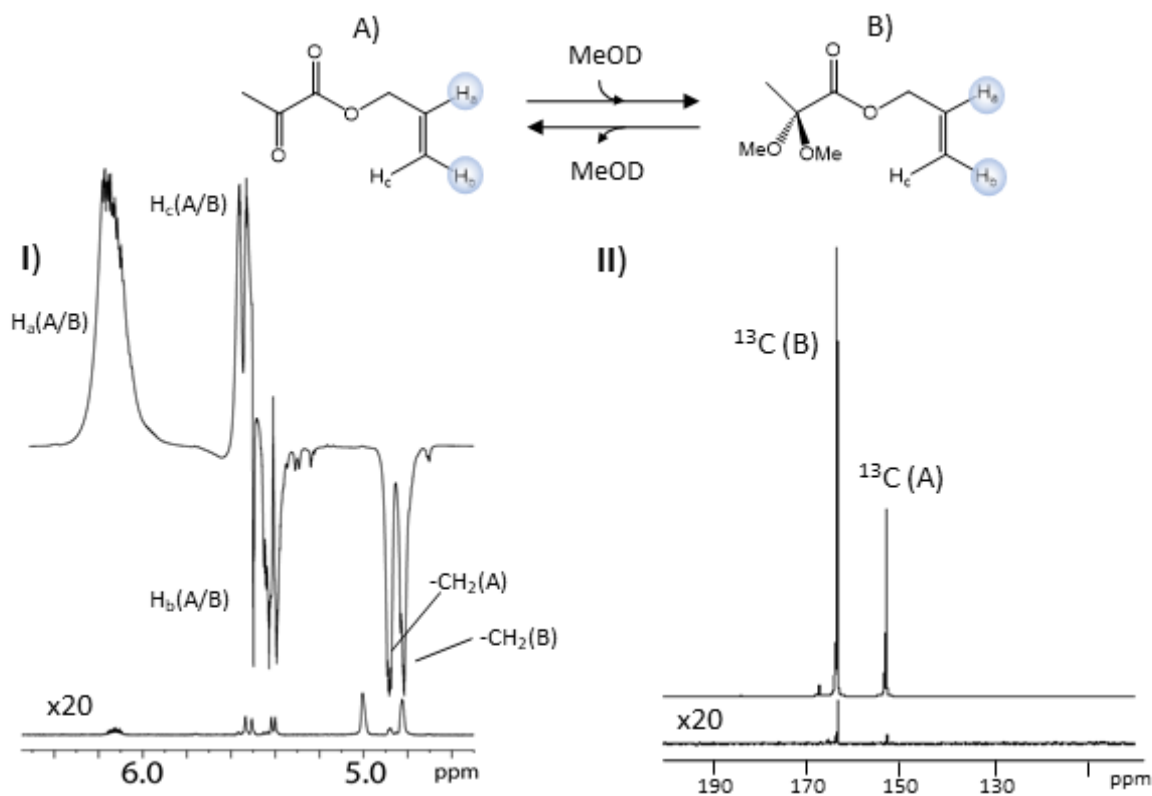

**Figure 2**. The two isoforms of the pyruvate ester (allyl-pyruvate): the oxo-form (A) is 15% while the ketal form (B) is 85% of the total. **I)** $^{1}$H hyperpolarized (upper) and thermal (lower) spectra of allyl-pyruvate after hydrogenation, using para-enriched hydrogen, in methanol-$d_4$ (ALTADENA experiments). **II)** $^{13}$C hyperpolarized (upper) and thermal (lower) spectra of allyl pyruvate obtained in the same solvent, after the application of magnetic field cycle for spin order transfer from parahydrogen protons to $^{13}$C. From the comparison between the hyperpolarized and thermal spectra, it can be easily noticed that the oxo- form (A) is significantly more polarized than the ketal form.

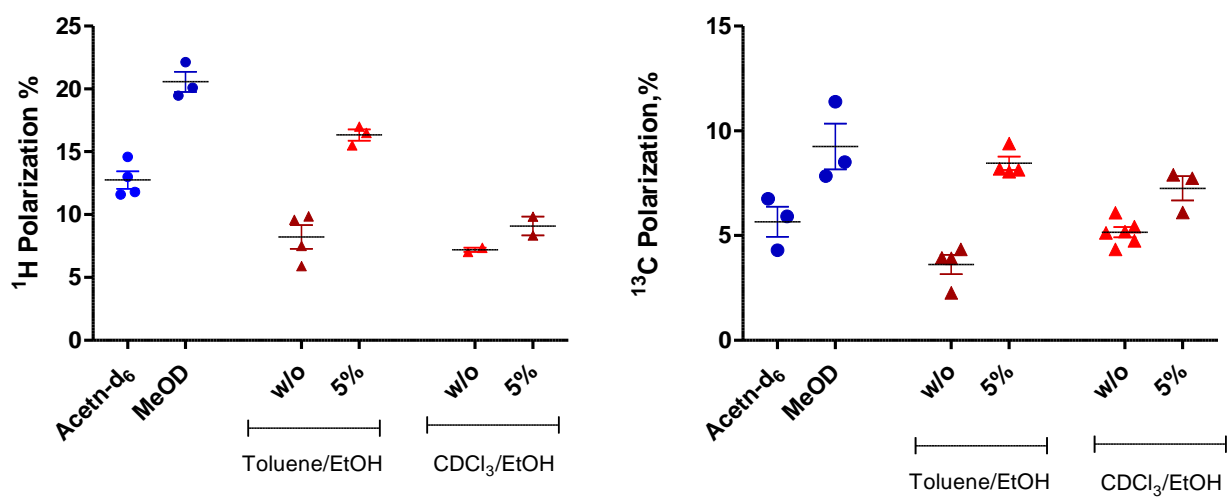

**Figure 3.** Hyperpolarization level on $^{1}$H and $^{13}$C-NMR signals obtained in different solvents. $^{1}$H hyperpolarization has been measured in ALTADENA experiments, while MFC (Magnetic Field Cycle) has been applied to obtain the spin order transfer from parahydrogen protons to $^{13}$C, in a different set of experiments.

| Solvent | $^1$H pol. | $^{13}$C pol. Allyl-[1-$^{13}$C]pyruvate | $^{13}$C pol. [1-$^{13}$C]Pyruvate | $^{13}$C pol. [1-$^{13}$C]Pyruvate Filtered |
|---|---|---|---|---|
| Acetone-d$_6$ | 12.7±1.4 | 5.7±1.2 | | |
| Methanol-d$_4$ | 20.6±1.4 | 9.2 ± 1.9 (17.5±0.9 oxo-form 6.4±0.2 ketalic-form) | | |
| Toluene-h8 | 6.7 ± 1% | 3.6 ± 0.9 | | |
| Toluene/EtOH (5%) | 16.3± 0.8 | 8.4 ± 0.6 | 4.1 ± 0.3 | 3.4 ± 0.6 |
| Chloroform-d | 7.2±0.2 | 5.2 ± 0.6 | 2.7 ± 0.6 | 2.2 ± 0.3 |
| Chloroform-d/EtOH (5%) | 9.1 ± 1 | 7.2 ± 1 6.2 ± 0.3* | 3.5 ± 0.5 * | |
| Chloroform/EtOH (20%) | | 8.5 ± 0-4** | 5.2 ± 0.2 ** | |

**Table 1**. Hyperpolarization level on $^1$H and $^{13}$C signals of parahydrogenated allyl-pyruvate. * $^{13}$C hyperpolarization values reported in ref.[44] and ** in [45].

### 3.1.3 Hydrolysis and phase extraction

For biological applications, hydrolysis of the allyl-ester and transfer of hyperpolarized pyruvate into the aqueous phase are the next steps. Phase transfer of the hyperpolarized product from the organic to the aqueous phase can be obtained, provided that the hydrogenation reaction is carried out in an organic hydrophobic solvent, therefore toluene/ethanol (5%) and pure chloroform have been tested.

Despite that high polarization has been observed in methanol, the aqueous solution of hyperpolarized pyruvate obtained using methanol as hydrogenation solvent cannot be applied for biological studies, being this solvent completely mixable with water. Analogously, special attention has to be devoted when the addition of ethanol to the organic phase is pursued to increase the efficiency of hydrogenation and hyperpolarization, as one has to keep in mind that high percentages of ethanol in the aqueous solution may not be compatible with the intended biological applications and in-vivo studies.

Hydrolysis of the ester is carried out by injecting an aliquot of a heated base solution in the NMR tube that contains the organic solution of the hydrogenation product. This step is followed, after few seconds (7-8s), by the addition of an acidic buffer solution in order to reach physiologically compatible pH values. The aqueous phase is withdrawn into a syringe and transferred into another sample tube for the acquisition of the NMR spectrum. All these passages, from the hydrolysis of the ester to the acquisition of the NMR spectrum, take a few tens of seconds (32-34s) (see video 1). Due to the ongoing relaxation processes, the measured hyperpolarization level of the $^{13}$C carboxylate signals resulted to be 4.1 ± 0.3 % when the hydrogenation solvent was the toluene-h$_8$/ethanol mixture, and 2.7 ± 0.6 % in the case of chloroform.

[1]H-NMR spectra of the aqueous solutions were acquired to determine the concentration of all the chemical species present in the solution of the hyperpolarized product(figure S3 and S4). Previously reported cytotoxicity studies on cells [18] have shown that a toxicity effect associated to the presence of the organic solvent (chloroform) is evident, when cells are incubated for 12-24h in the aqueous solution of products obtained by applying the hydrogenation/hydrolysis/phase extraction procedure above described. Other impurities derived from the hydrogenation reaction (allyl-alcohol, ethanol) do not have any effect on cells viability.

The final concentrations of chloroform and toluene, in the aqueous solution of pyruvate, are 30± 2 mM and 15±2 mM, respectively. These values are higher than those recommended by the Environmental Protection Agencies for water quality criteria[46] (0.59mM for chloroform and 10mM for toluene), therefore a purification step must be added in order to reduce as much as possible. To do that, filtration of the aqueous solution using a lipophilic resin was applied as the final step, at the end of the hyperpolarization procedure. Tenax® TA[47][48] is a common porous organic polymer used for the absorption of volatile organic compounds. The use of a PEEK filter to retain the resin allowed to avoid hyperpolarization losses during filtration, whereas it was observed that, if stainless steel filters are used, the hyperpolarization loss during filtration is significant (figure S5).

Filtration through the lipophilic resin allowed to completely remove toluene from the aqueous phase, while the chloroform concentration (0.5 ± 0.1 mM) resulted lower than the value recommended by EPA.

Unfortunately, the filtration step caused a decrease in the $^{13}$C polarization level that resulted to be 3.4 ± 0.6 % in the case of the toluene/ethanol (5% ethanol) solution was used and 2.2 ± 0.3 for chloroform as hydrogenation solvent, respectively. The decrease of the hyperpolarization level appears mainly due to the longer work-up of the solution. The filtration passage takes 18-20 seconds, therefore the time delay from the hydrolysis to the acquisition of the hyperpolarized spectrum becomes 50-52s. It can be reasonably expected that these polarization losses could be significantly reduced if an automatized procedure would be applied. Nevertheless, the $^{13}$C hyperpolarization obtained in the case of hydrogenation in toluene/ethanol is still sufficient for in-vivo metabolic investigations, since it is quite close to that reported for the first in-vivo study carried out using PHIP-SAH polarized pyruvate [44].

|  | [Pyruvate] | [allyl-alcohol] | EtOH | Solvent | Solvent (filtration) |
|---|---|---|---|---|---|
| Toluene/EtOH | 50 ±5 mM | 45± 5 mM | 210 ± 50 mM | 15 ± 3 mM | n.d. |
| CDCl$_3$ | 30 ± 5 mM | 8 ± 1 mM | - | 30 ± 2 mM | 0.5 ± 0.1 mM |

**Table 2.** Concentration of the different chemical species in the aqueous phase

### 4.1 Conclusions

The hyperpolarization level on the parahydrogenation products depends strongly on the solvent used. This is due not only to the solubility of hydrogen in a given solvent, but also to the binding interaction of the solvent molecules to the hydrogenation catalyst. Coordinating solvents such as acetone, methanol and ethanol can lead to high efficiency, in terms of hydrogenation speed and polarization level. The higher polarization level obtained in coordinating solvents can be explained considering that the displacement of the product molecule from the metal center is faster, thus limiting the singlet/triplet mixing on intermediates.

Reactions that were carried out using variable concentrations of ethanol in chloroform or toluene have clearly shown the dependence of the $^{13}$C polarization level and of the hydrogenation efficiency on the relative amount of the co-solvent. However, high percentages of the co-solvent are not compatible with biological applications, and have to be avoided in the preparation of samples for in-vivo studies.

In order to allow the phase-extraction of hyperpolarized pyruvate, chloroform and toluene/ethanol (5% v/v) have been used. In these solvents, $^{13}$C hyperpolarization on the allyl-ester are 5.2 ± 0.6% and 8.4 ± 0.6% respectively. After hydrolysis and phase extraction of sodium [1-$^{13}$C]pyruvate, the $^{13}$C polarization is 2.7 ± 0.6% when hydrogenation is carried out in chloroform and 4.1 ± 0.3% when the toluene/ethanol mixture is used. Although the solubility of these organic solvents in water is low, their concentration in the aqueous solution is non-neglectable. Filtration of the aqueous solution of the hyperpolarized product through a lipophilic resin (Tenax® TA) led to the complete removal of the organic solvents, still remaining a good hyperpolarization level almost unaffected, still sufficient for $^{13}$C MRI studies.

### Aknowledgements

This project has received funding from the European Union's Horizon 2020 research and innovation program under the Marie Skłodowska-Curie (Grant Agreement No. 766402) and the FETOPEN program (Grant agreement 858149, proposal acronym Alternatives to Gd). The Italian MIUR is also acknowledged for funding (PON Research and Innovation 2014-2020, CUP ARS01_00144, Novel molecular Imaging methods for the investigation of oncological and neurodegenerative diseases, MOLIM OncoBrain).

compounds (VOCs) on electrospun nanofibers with Tenax TA for potential application in sampling, PLoS One. 11 (2016) 1–14. doi:10.1371/journal.pone.0163388.

# Supplementary material

Magnetic field cycle.

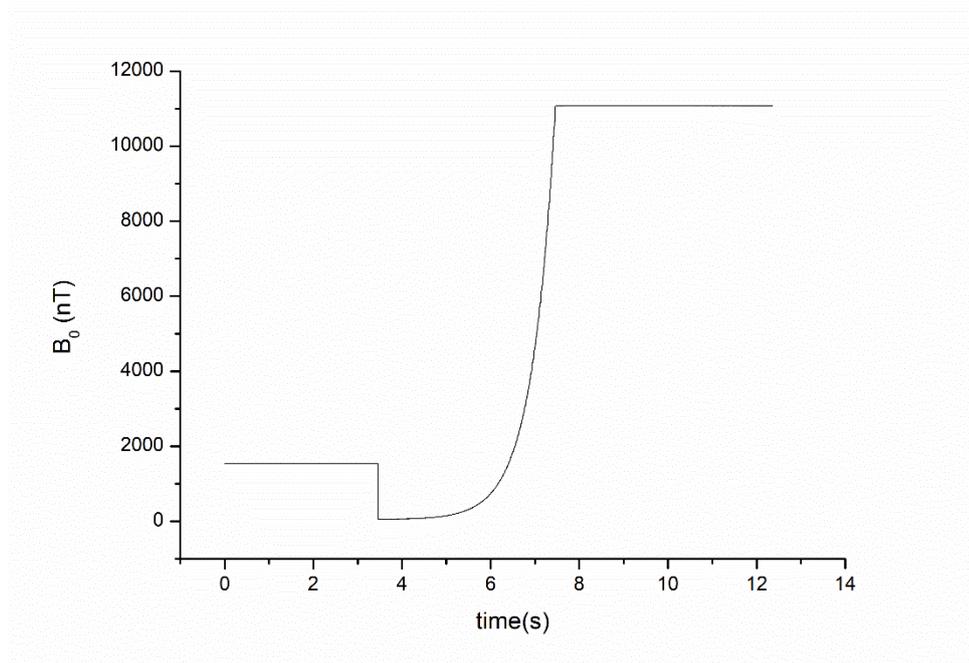

**Figure S1.** Magnetic field profile used for magnetic field cycle. The measurement has been carried out by means of a three axis fluxgate magnetometer (Mag-03, Bartington, Witney, UK).

|  | δ $^{31}$P (ppm) [Rh(cod)dppb]$^+$BF$_4$ | J$_{P-Rh}$ (Hz) [Rh(cod)dppb]$^+$BF$_4$ | δ $^{31}$P (ppm) [Rh(S$_2$)dppb]$^+$BF$_4$ S=Solvent | J$_{P-Rh}$ (Hz) [Rh(S$_2$)dppb]$^+$BF$_4$ S=Solvent |
|---|---|---|---|---|
| Acetone-d$_6$ | 24 ppm | 144 Hz | 50 ppm | 195 Hz |
| Methanol-d$_4$ | 24 ppm | 144 Hz | 53 ppm | 200 Hz |
| Toluene | 24 ppm | 145 Hz | 41 ppm | 201 Hz |
| Toluene/Ethanol (5% v/v) | 24 ppm | 144 Hz | 41 ppm | 201 Hz |
| Ethanol-d$_4$ | 24 ppm | 144 Hz | 52 ppm | 200 Hz |
| CDCl$_3$ | 24 ppm | 144 Hz | 29 ppm; 46 ppm | - |
| CDCl$_3$/Ethanol (5% v/v) | 24 ppm | 144 Hz | 29 ppm; 46 ppm | - |

**Table S1**. $^{31}$P-NMR signals of the catalyst in different solvents before activation ([Rh(cod)dppb]$^+$BF$_4$) and after hydrogenation of the coordinated diene. In coordinating solvents, the complex ([Rh(S$_2$)dppb]$^+$BF$_4$ is formed and the effect on the $^{31}$P-NMR signal can be observed clearly. Chloroform is non-coordinating and other species (dimers) are formed.

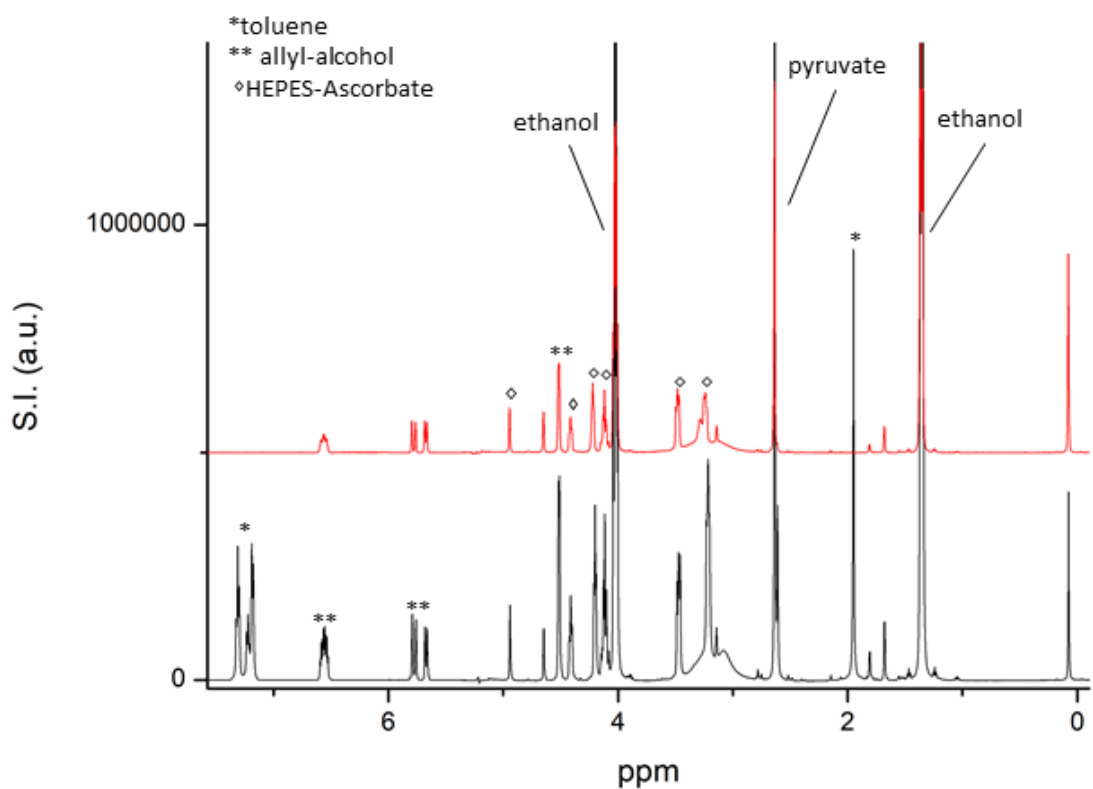

**Figure S2.** Black line: $^1$H-NMR of aqueous solutions of sodium[1-$^{13}$C]pyruvate obtained after hydrolysis following to the hydrogenation carried out in the toluene/ethanol solution. Red line: $^1$H-NMR of aqueous solutions of sodium[1-$^{13}$C]pyruvate obtained after hydrolysis (as in the upper spectrum) and filtration through the Tenax column (see main text, Materials and methods section, for the experimental details).

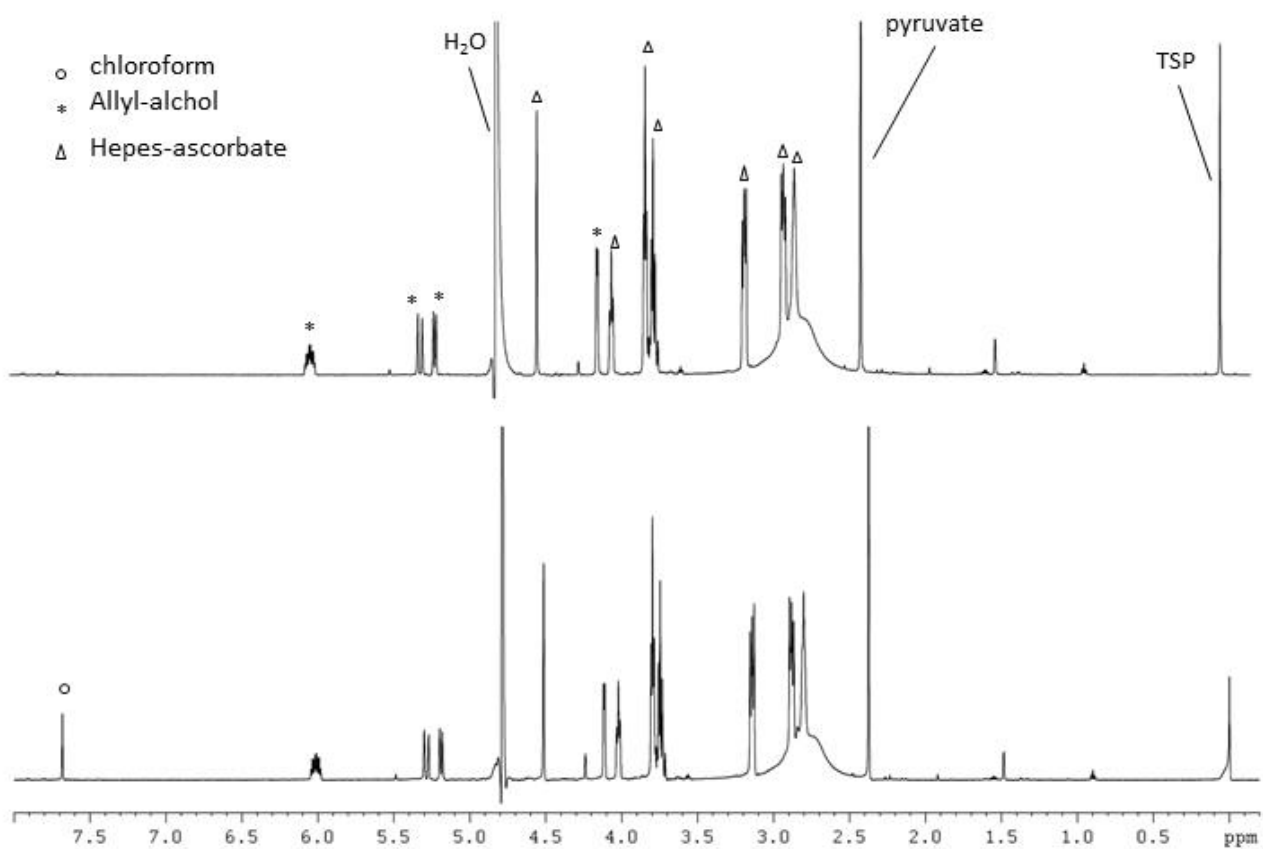

**Figure S3.** Lower spectrum: $^1$H-NMR spectrum of the aqueous solution of sodium [1-$^{13}$C]pyruvate obtained from hydrolysis of the ester (propargyl[1-$^{13}$C]pyruvate) hydrogenated in chloroform-d. Upper spectrum: $^1$H-NMR spectrum of the aqueous solution of sodium [1-$^{13}$C]pyruvate obtained from hydrolysis of the ester (as in the upper spectrum) and filtration through the Tenax resin (100 µl in a syringe, retained by a PEEK filter, as reported in the main text, Materials and Methods).

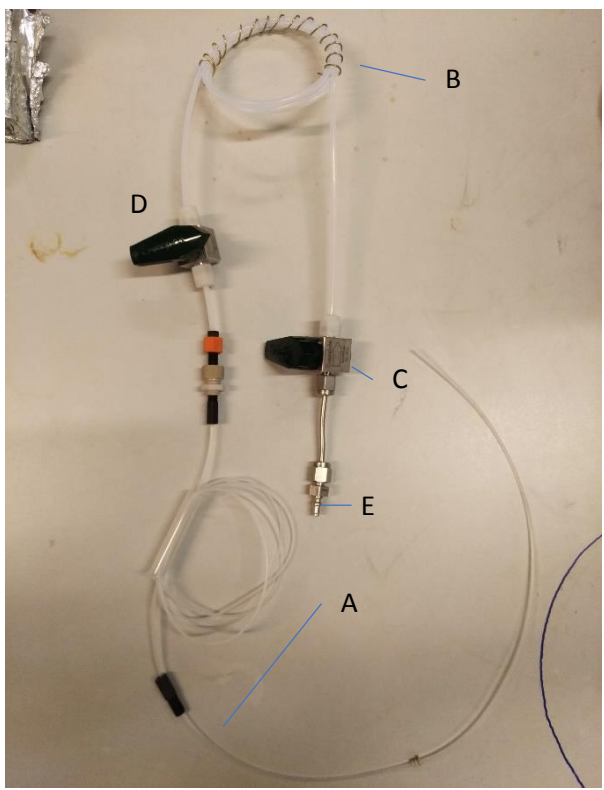

**Figure S4.** Setup for hydrolysis: the base solution is loaded in tube A; Ar is connected to E, while D is kept closed, and tube B is pressurized using argon (1.5bar); valve D is closed. The setup is hold vertically and tube A is plunged in a hot water bath. For the injection of the aqueous base into the organic solution (hydrolysis step), the tip of tube A is placed a few mm above the organic solution of the product, then valve D is opened.

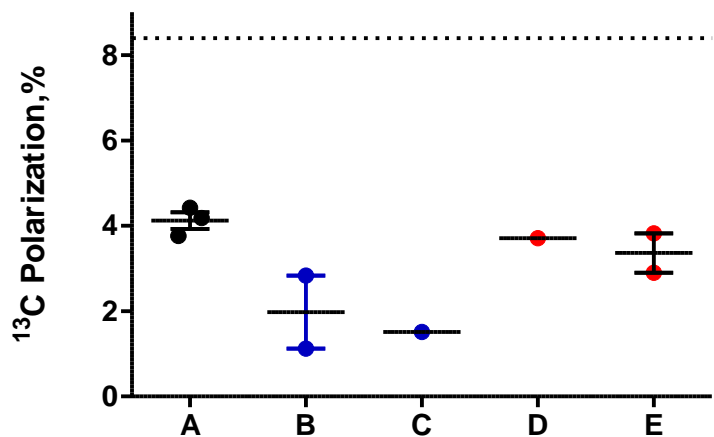

**Figure S5.** $^{13}$C polarization on [1-$^{13}$C]pyruvate in the aqueous phase obtained after phase extraction from toluene/ethanol (ethanol 5% v/v). A) after hydrolysis, without filtration; B) after hydrolysis and passage through a SS filter (Idex Fluidics, Frit in Ferrule 1/16" PEEK Super Flangeless, SS Frit 2um); C) as in B, with Tenax resin; D) hydrolysis and passage through a PEEK filter (Idex Fluidics, Frit in Ferrule 1/16" PEEK Super Flangeless, PEEK Frit 2um); E) as in E, with Tenax resin added.